\documentclass[prd,superscriptaddress,floatfix,amsfonts,amssymb,amsmath,showpacs,twocolumn]{revtex4}
\usepackage{bm}
\usepackage{amsfonts}
\usepackage{latexsym}
\usepackage[latin1]{inputenc}
\usepackage{graphicx}
\usepackage{amsmath}
\usepackage{palatino}
\usepackage{mathpazo}
\usepackage{textcomp}
\usepackage{booktabs}
\usepackage{dcolumn}
\usepackage{hyperref}
\hypersetup{colorlinks,citecolor=blue}
\usepackage{amsmath}
\usepackage{xcolor}
\usepackage{mathtools}
\usepackage{multirow}
\usepackage{float}

\allowdisplaybreaks[1]

\begin{document}
\color{red}
	
\title{Redshift Drift in $f(R,T)$ Gravity}

\author{Snehasish Bhattacharjee}
\email{snehasish.bhattacharjee.666@gmail.com} 
\affiliation{Department of Astronomy, Osmania University, Hyderabad-500007,
India}
\author{P.K. Sahoo}
 \email{pksahoo@hyderabad.bits-pilani.ac.in}
\affiliation{ Department of Mathematics, Birla Institute of Technology and Science-Pilani,\\ Hyderabad Campus,Hyderabad-500078, India}

\begin{abstract}

Redshift drift refers to the phenomena that redshift of cosmic objects is a function of time. Measurement of redshift drift is of fundamental importance in physical cosmology and can be utilized to distinguish different cosmological models. Redshift drift can be expressed in two distinct methods. The first method is related to cosmography, where the Redshift drift is given as a series expansion of cosmological parameters, while the second method is written as a function of Hubble parameter and its time derivatives which ultimately involve field equations of a chosen theory of gravity. By equating corresponding terms from both the series, the model parameter(s) of any modified theory of gravity can be constrained. The present note aims at constraining the model parameter $\zeta$ of $f(R,T)$ gravity theory where $f(R,T)= R + \zeta T$. By equating linear terms in redshift $z$ from both the series, we constrain $\zeta$ in the range $-0.51 \kappa^{2} \lesssim \zeta \lesssim -0.47 \kappa^{2}$, where $\kappa^{2}=\frac{8 \pi G}{c^{4}}$.    
\end{abstract}



\maketitle

\section{Introduction}\label{I}

Redshift drift (hereafter `RD') refers to the phenomena that redshift of cosmic objects is a function of time and was first proposed in \cite{sandage,vit}. In 1998 Loeb \cite{loeb} demonstrated a possible way to detect this effect by employing the Lyman-$\alpha$ forest and is known as Sandage-Loeb test.\\
Conceptually, a measurement of RD is of fundamental importance in physical cosmology owing to the fact that it is a model-independent examination of the cosmic expansion, without a priori presumption on geometry, large scale clustering or nature of  gravity \cite{alves}. However, determining this effect requires high sensitivity which is nearly impossible with most astronomical facilities at the present time. In order to detect this effect, a high-resolution spectrograph (ELT- HIRES) was appended to the Extremely Large Telescope (ELT) \cite{liske2014}. A comprehensive decade long study was conducted on high-redshift objects in  \cite{liske2008} and a list of possible targets were reported there. The paramount of these observations for bridling cosmological models were reported in \cite{cora,lake,balbi,moreas,geng}.\\
The current accelerated expansion of the cosmos can be elegantly described in various modified gravity theories without engaging the idea of dark energy (DE). Out of these, $f(R,T)$ gravity is one of the most frequently studied modified gravity theory. First introduced in the literature \cite{harko}, the theory is formulated by replacing the Ricci scalar $R$ in the Einstein-Hilbert action by a combined function of $R$ and trace of energy momentum tensor $T$. Since the function can be arbitrarily chosen with suitable model parameter(s), it (they) can be tuned to fit the observations gracefully. Viable csomological models in $f(R,T)$ gravity have been studied in \cite{cm14,cm15,cm16,cm17,cm18,ph1,z1,z2} without employing DE, while in \cite{dm} the authors modeled a flat galactic rotation curve without dark matter using $f(R,T)$ gravitation. $f(R,T)$ gravity has also been applied to other cosmological scenarios such as bouncing cosmology \cite{bhatta,PDU}, baryogenesis \cite{baryo,baryo2}, Big-Bang nucleosynthesis \cite{bang}, growth rate of matter fluctuations \cite{growth}, varying speed of light scenarios \cite{physical}, wormholes \cite{in26,yousaf} and irregular energy density \cite{density}. \\
The present note aims at utilizing the idea of RD in constraining the model parameter ($\zeta$) to a very high accuracy.\\
RD can be expressed in two distinct methods. The first method is related to cosmography, where the RD is given as a series expansion of cosmological parameters \textit{i.e,} Hubble parameter, deceleration prameter, jerk parameter, snap parameter, pop parameter, etc \cite{j1,j2,j3}, while in the second method, RD is written as a function of Hubble parameter and its higher order time derivatives. In the latter method, Hubble parameter is written in terms of field equations of a chosen theory of gravity. By equating the coefficients of $z^{i}$, where $i=1,2...\infty$, from these two series expansions, the model parameter(s) of any modified gravity can be constrained.\\ 
The manuscript is organized as follows: In Section \ref{sec2} we present an overview of RD in cosmographical and dynamical approaches. In Section \ref{sec3} we briefly discuss Cosmography in $\Lambda$CDM model. In Section \ref{sec4} we present an overview of $f(R,T)$ modified gravity and obtain the expression of Hubble parameter. In Section \ref{sec5} we utilize the expression of Hubble parameter and equate the corresponding linear terms in $z$ from both the series expansions and finally constrain the value of $\zeta$. Finally Section \ref{sec6} is devoted to conclusions and discussions.

\section{Redshift Drift}\label{sec2}

Redshift of an astronomical source emitting a photon at instant $t$ and observed at instant $t_{0}$, is given by:
\begin{equation}
\frac{a(t_{0})}{a(t)} = 1 + z
\end{equation}
Owing to the non-uniform expansion of the cosmos, the redshift $z$ is time dependent $z(t)$. As a result, a second photon emitted by the same object at different instant $t + \bigtriangleup t=t^{'}$, will correspond to a redshift $z(t^{'})$. This phenomena is refereed to as \textit{redshift drift}, which can be approximated as \cite{loeb}
\begin{equation}
\frac{\dot{a(t_{0})}-\dot{a(t)}}{a(t)} = \frac{\bigtriangleup z} {\bigtriangleup t_{0}} =( 1 + z) H_{0} - H(z)
\end{equation}
where $\bigtriangleup t_{0}$ correspond to the time interval between two observed photons. Remarkably, this observable neither depends on how we define a standard ruler nor on the intrinsic luminosity or any other characteristics of the cosmic object.\\
Two different series expansions expressing RD shall now be presented involving this observable as a function of $z$. The first series involves kinematic and geometric attributes of the metric (cosmographic approach), while the second series is  represented in terms of the Hubble parameter and its time derivatives which is purely dependent on the chosen gravitational theory (dynamical approach).\\
\begin{itemize}

\item \textbf{Cosmographical Method}:\\
The kinematic and geometric attributes of the metric are designated by scale factor and its time derivatives around $t_{0}$. This reads
\begin{eqnarray}
\left(\frac{\dot{a}}{a} \right) \bigg|_{t0}\equiv H_{0},\  \frac{-1}{H_{0}^{2}}\left(\frac{\ddot{a}}{a} \right)\bigg|_{t0}\equiv q_{0},\  \frac{-1}{H_{0}^{3}}\left(\frac{\dddot{a}}{a} \right)\bigg|_{t0}\equiv j_{0}
\end{eqnarray}
where overhead dots represent time derivatives and $'0'$ denote the evaluation of these quantities at the present epoch ($z=0$). This approach aims at estimating RD utilizing these quantities.\\
Expanding $H(z)$ in terms of these parameters and coupled with simple algebra, RD can be expressed as 
\begin{equation}\label{1}
\frac{\bigtriangleup z} {t_{0}} = -q_{0}H_{0}z - \frac{1}{2} \left( j_{0} - q_{0}^{2} \right)H_{0} z^{2} + \mathcal{O} (z^{3})  
\end{equation}
Where $\frac{\bigtriangleup z} {t_{0}}$ is a function of redshift ($z$). It is reported that $\frac{\bigtriangleup z} {t_{0}}$ is positive for $0\leq z \lesssim 2$ with maximum value at $z\sim 1.2$. For $z\gtrsim 2$, $\frac{\bigtriangleup z} {t_{0}} < 0$ and decreases with increasing redshift \cite{bole}.\\
Since $H_{0}^{-1}\simeq t_{0}$, where $t_{0}$ represent current age of the universe, we can write \eqref{1} as
\begin{equation}\label{2}
\frac{\bigtriangleup z} {t_{0}} = -\frac{q_{0}}{t_{0}}z - \frac{1}{2} \left( j_{0} - q_{0}^{2} \right)\frac{1}{t_{0}} z^{2} + \mathcal{O} (z^{3})  
\end{equation}
\item \textbf{Dynamical Method}:\\
In this approach, the RD is written as a series expansion of Hubble parameter $H(z)$ and its time derivatives. Thence, the gravitational field equations enters the scenario through the Hubble parameter which depends on the chosen theory of gravity. 
The expression of RD for the dynamical method reads
\begin{equation}\label{3}
\frac{\bigtriangleup z} {t_{0}} =\left(\frac{\dot{H_{0}}}{H_{0}} + H_{0} \right)z + \left(\frac{\dot{H_{0}}}{H_{0}} - \frac{\dot{H_{0}}^{2}}{H_{0}^{3}} + \frac{\ddot{H_{0}}}{H_{0}^{2}} \right)\frac{z^{2}}{2}  + \mathcal{O} (z^{3})
\end{equation}
\end{itemize}

\section{Cosmography in $\Lambda$CDM model}\label{sec3}
Cosmography is a convenient perspective to distinguish and study cosmological models framed completely on Cosmological Principle \cite{33}, and provide useful insights of the cosmic expansion and evolution of cosmological parameters \cite{34,35,36}.\\
By comparing the coefficients of different powers in $z$, we obtain an infinite number of constrained equations. The first elements of the series reads:
\begin{equation}\label{4}
q_{0}= -\Omega_{\Lambda 0} + \frac{1}{2}\Omega_{m 0}
\end{equation}
\begin{equation}\label{5}
j_{0}=\Omega^{2}_{\Lambda 0} - \Omega_{m 0}\Omega_{\Lambda 0} - 3\Omega_{m 0} +\frac{5}{2}\Omega^{2}_{m 0}
\end{equation}
Hence from the knowledge of density parameters $\Omega_{\Lambda 0}$ and $\Omega_{m 0}$ we can estimate $q_{0}$ and $j_{0}$. Plugging $\Omega_{m 0} =0.315 \pm 0.007$ and $\Omega_{\Lambda 0}=0.6847 \pm 0.0073$ \cite{planck}, we obtain
\begin{equation}\label{6}
q_{0} \simeq -0.5271 \pm 0.008
\end{equation}
\begin{equation}\label{7}
j_{0} \simeq -0.4439  \pm 0.0165
\end{equation}
the ascertained values agrees well with values deduced from other observations \cite{38,39,40}.
\section{Overview of $f(R,T)$ Gravity}\label{sec4}
For the $f(R,T)$ modified gravity, action is given by 

\begin{equation}\label{8}
\mathcal{S}_{(R,T)}=\frac{1}{2\kappa^{2}}\int  d^{4}x \sqrt{-g}\left[ f(R,T)+\mathcal{L}_{m} \right]   
\end{equation}
where $\kappa^{2}=\frac{8 \pi G}{c^{4}}$, $\mathcal{L}_m=-p$ denote matter Lagrangian and $p$ is the cosmic pressure.\\
Variation of \eqref{8} with metric ($g_{ij}$) yields the $f(R,T)$ gravity field equation as
\begin{equation}
\kappa^{2}T_{ij}-f^{1}_{,T}(R,T)(T_{ij} + \Theta_{ij})=\Pi_{ij}f^{1}_{,R}(R,T)+f^{1}_{,R}(R,T)R_{ij} -\frac{1}{2}g_{ij}f(R,T)
\end{equation}
where, $T_{ij}$ represent energy momentum tensor of a perfect fluid and reads
\begin{equation}
( \rho + p)u_{i}u_{j}- pg_{ij}=T_{ij}
\end{equation}
where $\rho$ represent matter density.
\begin{equation}
\Pi_{ij}= g_{ij}\square-\nabla_{i} \nabla_{j}
\end{equation}
\begin{equation}
\Theta_{ij}\equiv g^{\mu \nu}\frac{\delta T_{\mu \nu}}{\delta g^{ij}}
\end{equation}
The notation $f^{i}_{,X}\equiv \frac{d^{i}f}{d X^{i}}$.
We set $f(R,T) = R + \zeta T$. Modified Friedman equations for a flat FLRW spacetime with ($-$,$+$,$+$,$+$) metric signature, reads
\begin{equation}
3H^{2}= \left[ \frac{1}{2}\left(  3-\omega\right)\zeta + \kappa^{2} \right] \rho
\end{equation} 
\begin{equation}
-2\dot{H}-3H^{2} = \left[\frac{1}{2}\left( 3 \omega-1\right)\zeta + \kappa^{2} \omega  \right] \rho
\end{equation}
where $\omega = p/ \rho$ represent equation of state (EoS) parameter.
By eliminating $\rho$ and solving for $H$, we obtain 
\begin{equation}\label{9}
H=\frac{\theta}{t}
\end{equation}
Thence, time derivative of Hubble parameter ($\dot{H}$) reads
\begin{equation}
\dot{H}= -\frac{\theta}{t^{2}}
\end{equation}
where, 
\begin{equation}\label{10}
\theta = \frac{\left( 3 - \omega\right)\zeta + 2\kappa^{2} }{3\left(1 + \omega\right)\left(\kappa^{2}  + \zeta\right)   }
\end{equation}

\section{Constraining $\zeta$}\label{sec5}

Two distinct methods (cosmographical and dynamical) were presented in Section \ref{sec2} to ascertain the RD. As represented by the $\Lambda$CDM cosmological model, equating the coefficients of $z^{i}$ from both the series, results in a concatenation of constraint equations which can then be solved. In case of modified gravity theories, these equations relate model parameters of a chosen theory of gravity to the cosmological parameters ($H_{0}$, $q_{0}$, $j_{0}$, etc). Since the values of these cosmological parameters are well known from various observations, they can be employed to set bounds on the model parameter(s). Here, we aim at evaluating $\zeta$ for the $f(R,T)$ gravity model $f(R,T)= R + \zeta T$. This is one of the most widely studied $f(R,T)$ gravity model (check, for instance, \cite{harko,m23,m25,m31}).\\
Equating the coefficients of $z^{i}$ where, $i=1$, from \eqref{2} and \eqref{3}, we obtain
\begin{equation}\label{11}
\left(\frac{\theta}{t_{0}} - \frac{1}{t_{0}}\right) = -\left(\frac{1}{t_{0}} \right)q_{0} 
\end{equation} 
Substituting \eqref{10} in \eqref{11} we obtain 
\begin{equation}\label{12}
\frac{(3-\omega)\zeta + 2 \kappa^{2}}{(\kappa^{2}+\zeta)}=3(1-q)(1+\omega)
\end{equation}

\textbf{Observational Constraint}: Cosmological observations indicate $q_{0}=-0.57 ^{+0.10}_{-0.08}$ \cite{super} \& $\omega_{0}=-1.03 \pm 0.03$ \cite{planck}. Substituting these values in \eqref{12}, we found $\zeta$ to lie in the range $-0.51 \kappa^{2} \lesssim \zeta \lesssim -0.47 \kappa^{2}$. The result indicates positive values of $\zeta$ are not permissible. 

\section{Conclusions}\label{sec6}

Modified gravity theories are viable alternatives to GR in explaining the current cosmic acceleration without dark energy. However, attempts must be made to put constraints on the model parameters of these extended theories of gravity to understand their efficiency and applicability. In the present manuscript, we seek to constrain the model parameter $\zeta$ of one of the most widely studied $f(R,T)$ gravity model of the form $f(R,T) = R + \zeta T$. We achieved this by writing RD in two distinct series expansions and then equating corresponding terms from both the series. The first series is written in terms of cosmological parameters ($H_{0}$, $q_{0}$, $j_{0}$, etc). These parameters can be estimated from various cosmological observations. The second series is written as a function of Hubble parameter and its time derivatives which ultimately involve field equations of $f(R,T)$ gravity. By equating the linear terms in redshift $z$ from both the series, we constrain $\zeta$ in the range $-0.51 \kappa^{2}\lesssim \zeta \lesssim -0.47 \kappa^{2}$, where $\kappa^{2}=\frac{8 \pi G}{c^{4}}$.     \\
Our result makes it clear-cut that higher order $T$ terms in the action would yield minute changes in cosmological models.\\
As a final note would add that the limits imposed on $\zeta$ has no dependency on the actual computation of the RD. Constraints yielded from theoretical estimations should also be considered to decide if  $f(R,T)$ gravity models provide satisfactory representation of the cosmos.

\acknowledgments PKS acknowledges DST, New Delhi, India for providing facilities through DST-FIST lab, Department of Mathematics, BITS-Pilani, Hyderabad Campus where a part of this work was done. We are very much grateful to the honourable referee and the editor for the illuminating suggestions that have significantly improved our work in terms of research quality and presentation.

\end{document}